\documentclass[conference]{IEEEtran}
\renewcommand\IEEEkeywordsname{Keywords}
\usepackage{cite}
\usepackage{amsmath,amssymb,amsfonts}
\usepackage{algorithmic}
\usepackage{graphicx}
\usepackage{textcomp}
\usepackage{xcolor}
\usepackage[hidelinks]{hyperref}
\def\BibTeX{{\rm B\kern-.05em{\sc i\kern-.025em b}\kern-.08em
    T\kern-.1667em\lower.7ex\hbox{E}\kern-.125emX}}

\usepackage{amsmath}
\usepackage{graphicx}
\usepackage{multirow}

\makeatletter

\def\ps@IEEEtitlepagestyle{%
  \def\@oddfoot{\mycopyrightnotice}%
  \def\@evenfoot{}%
}
\def\mycopyrightnotice{%
  {\footnotesize 978-1-6654-9106–8/22/\$31.00~\copyright~2022 IEEE\hfill}
  \gdef\mycopyrightnotice{}
}

\@ifundefined{showcaptionsetup}{}{
 \PassOptionsToPackage{caption=false}{subfig}}
\usepackage{subfig}
\makeatother

\usepackage{eso-pic}
\newcommand\AtPageUpperMyright[1]{\AtPageUpperLeft{%
 \put(\LenToUnit{0.5\paperwidth},\LenToUnit{-1cm}){%
     \parbox{0.5\textwidth}{\raggedleft\fontsize{9}{11}\selectfont #1}}%
 }}%
\newcommand{\conf}[1]{%
\AddToShipoutPictureBG*{%
\AtPageUpperMyright{#1}
}
}

\conf{
\textit{Proc. of the 8th International Conference on Engineering and Emerging Technologies (ICEET) \\
27- 28 October 2022, Kuala Lumpur, Malaysia}
} 
\begin{document}

\title{PCONet: A Convolutional Neural Network Architecture to Detect Polycystic Ovary Syndrome (PCOS) from Ovarian Ultrasound Images\\
}

\author{\IEEEauthorblockN{A.K.M. Salman Hosain}
\IEEEauthorblockA{\textit{Computer Science and Engineering} \\
\textit{Brac Universiy}\\
Dhaka, Bangladesh \\
akm.salman.hosain@g.bracu.ac.bd}
\and
\IEEEauthorblockN{Md Humaion Kabir Mehedi}
\IEEEauthorblockA{\textit{Computer Science and Engineering} \\
\textit{Brac Universiy}\\
Dhaka, Bangladesh \\
humaion.kabir.mehedi@g.bracu.ac.bd}
\and
\IEEEauthorblockN{Irteza Enan Kabir}
\IEEEauthorblockA{\textit{Electrical and Computer Engineering} \\
\textit{University of Rochester}\\
NY, United States \\
irtezaenan@gmail.com}
}

\maketitle

\begin{abstract}
Polycystic Ovary Syndrome (PCOS) is an endrocrinological dysfunction prevalent among women of reproductive age. PCOS is a combination of syndromes caused by an excess of androgens — a group of sex hormones — in women. Syndromes including acne, alopecia, hirsutism, hyperandrogenaemia, oligo-ovulation, etc. are caused by PCOS. It is also a major cause of female infertility. An estimated 15\% of reproductive-aged women are affected by PCOS globally. The necessity of detecting PCOS early due to the severity of its deleterious effects cannot be overstated. In this paper, we have developed PCONet - a Convolutional Neural Network (CNN) - to detect polycistic ovary from ovarian ultrasound images. We have also fine tuned InceptionV3 - a pretrained convolutional neural network of 45 layers - by utilizing the transfer learning method to classify polcystic ovarian ultrasound images. We have compared these two models on various quantitative performance evaluation parameters and demonstrated that PCONet is the superior one among these two with an accuracy of 98.12\%, whereas the fine tuned InceptionV3 showcased an accuracy of 96.56\% on test images.

\end{abstract}



\section{Introduction}
Hyperandrogenism and polycystic ovary morphology (PCOM) are two of the defining characteristics of Polycystic Ovary Syndrome, also known as PCOS- ovarian dysfunction~\cite{laven2002new}. Stein and Leventhal~\cite{stein1935amenorrhea} first defined PCOS in 1935 as a combined syndrome of hirsutism, chronic anovulation, amenorrhea, infertility, and enlarged cysts in the ovaries. But nevertheless, the World Health Organization (WHO) did not include `E28.2 Polycystic Ovarian Syndrome' in International Classification of Diseases, 10th revision, until 1990~\cite{world1992icd}.

Menstrual abnormalities, symptoms of an excess of androgen, and obesity are all possible clinical manifestations of this condition. There is also a correlation between PCOS and an escalated type 2 diabetes risk~\cite{rotterdam2004revised}. PCOS is defined as the prevalent reason for oligo-ovulatory infertility, which affects around 4\%-20\% reproductive aged women worldwide~\cite{deswal2020prevalence}. The risk of developing PCOS in patients with a family history of PCOS were 24\% and 32\%, respectively, but untreated premenopausal women were at higher risk of developing PCOS~\cite{kahsar2001prevalence}.

PCOS is a heterogeneous dysfunction which is a combination of various clinical symptoms, including hyperandrogenism (hirsutism, hyperandrogenaemia), dysfuntion of the ovary (oligo-ovulation, PCOM)~\cite{azziz2009androgen}. The most prevalent PCOS categorization is the Rotterdam definition, which is supported by the majority of scientific associations and health agencies at present~\cite{legro2013diagnosis}. According to the definition~\cite{rotterdam2004revised}, PCOS is proposed to be diagnosable in a patient exhibiting no less than two out of the three symptoms, which are:

\begin{itemize}
    \item Hyperandrogenemia 
    \item PCOM
    \item Ovulation disorder
\end{itemize}

However, the 2006 Androgen Excess and PCOS Society (AE–PCOS) necessitates hyperandrogenemia, ovulation disorder, and/or polycystic ovary morphology as symptoms to be defined as PCOS~\cite{azziz2006criteria}.

The criterion which are used to define PCOS has independent clinical effects~\cite{escobar2018polycystic}, such as: excessive androgen possibly causing cutaneous manifestations~\cite{azziz2009androgen}; ovulation disorder and chronic oligomenorrhoea causing endometrial hyperplasia, infertility, and carcinoma~\cite{azziz2009androgen}; and PCOM possibly causing an increased risk of ovarian hyperstimulation syndrome (OHSS)- a serious complication of ovulation induction~\cite{jayaprakasan2012prediction}. Generally, the severity of an individual patient's phenotype can be roughly correlated to the number of criteria that she satisfies for having PCOS~\cite{escobar2014menstrual}.

\begin{figure}[htp]
    \centering
    \includegraphics[width=8cm]{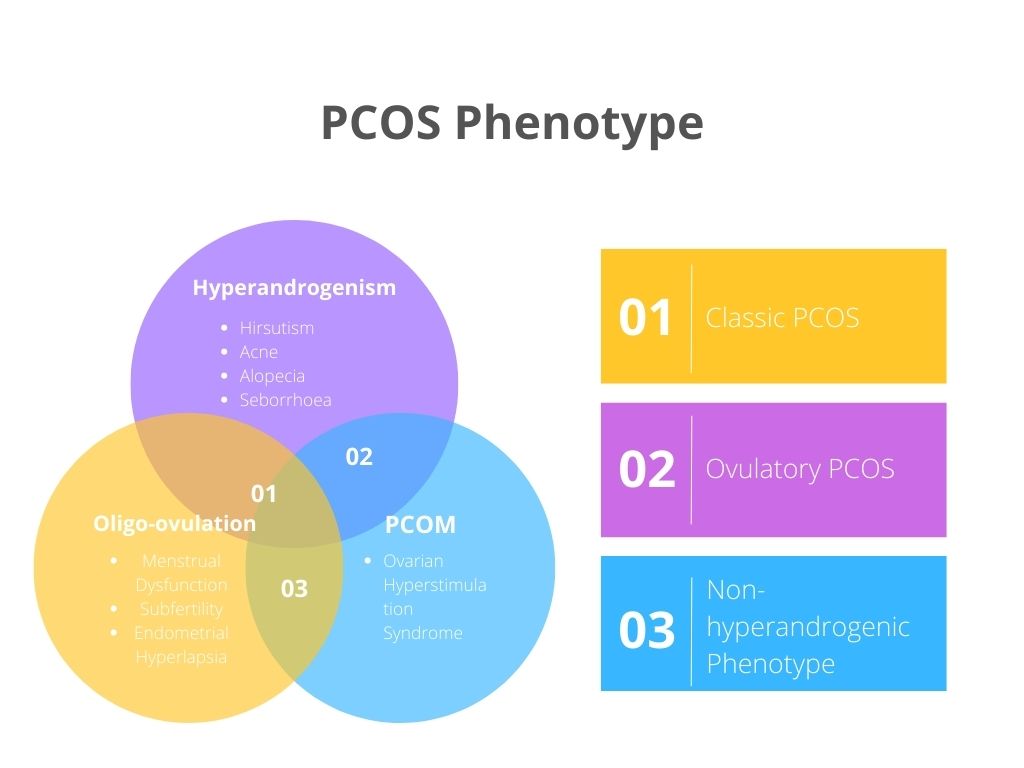}
    \caption{PCOSPhenotypes~\cite{escobar2014menstrual}}
    \label{fig:phenotype}
\end{figure}

The classic PCOS phenotype is the most severe clinical manifestation of PCOS, which, as shown in Fig. \ref{fig:phenotype}, regardless of the presence of PCOM, manifests with both hyperandrogenism and oligo-ovulation. The second most severe penotype is ovulatory PCOS, which manifests hyperandrogenism and PCOM. The least severe phenotype is the non-hyperandrogenic PCOS- consisting of oligo-ovulation and PCOM~\cite{escobar2014menstrual}- which, according to the AE–PCOS statement, is not to be considered PCOS~\cite{azziz2009androgen}.

PCOS is usually linked to abdominal adiposity. It also causes obesity, insulin resistance, cardiovascular risk factors, and metabolic disorders. PCOS is also associated with decreased fertility in women~\cite{escobar2018polycystic}. Presence of an ovarian volume of 10 ml and/or 25 follicles per ovary when using 8Mhz transducer frequencies is required to be defined as PCOM~\cite{dewailly2014definition}. Detectiing PCOS as early as possible is vital to start treatment. Use of machine learning approaches can be an alternative to detecting PCOS in traditional approach - which is often too time consuming.

Our contributions in this paper- 
\begin{itemize}
 \item From reviewing previous publications on PCOS detection using deep learning, we have noticed that works on PCOS detection from ultrasound images were significantly lower than classification based on data tables with numerical and categorical data on PCOS.

 \item We have developed a convolutional neural network (CNN) model, named PCONet, to detect ovarian cysts from ovarian ultrasound images. Our model can detect ovarian cysts with 98.12\% accuracy, which is higher than the pretrained~\cite{hasib2022covid} model InceptionV3 in classifying the ultrasound images of our test dataset.

 \item We have fine tuned and customized a convolutional neural network (CNN) model, InceptionV3, according to our dataset to detect ovarian cysts from ovary ultrasound images. This model showcased an accuracy of 96.56\%.

 \item We have built an ovarian ultrasound image dataset to test our models by collecting images from Google and labeling them with the help of two certified physicians. We used this discrete dataset as test set to ensure an unbiased performance evaluation of our models.
\end{itemize}


\section{Related Works}

Using k-nearest neighbor (KNN), decision tree classifier, naive bayes, logistic regression (LR), and support vector machine (SVM) Chauhan et al.~\cite{chauhan2021comparative} attempted to diagnose PCOS and discovered that Decision Tree was the most successful method in their particular circumstance.

Support vector machine, classification and regression trees (CART), naive bayes classification, random forest, and logistic regression were the classifiers that Hassan and Mirza~\cite{hassan2020comparative} applied in order to identify PCOS from the clinical data table of patients. The random forest classifier had the highest accuracy, which was 96\%.

Kumari~\cite{kumari2021classification} classified PCOS based on an image dataset using DenseNet-121, VGG-19, DenseNet-121, InceptionV3, and ResNet-50. VGG-19 had the highest accuracy out of the four, coming in at 70 percent, according to her findings. The author of this study overcame the constraints imposed by the dataset by utilizing the architecture of Generative Adversarial Networks (GANs) and augmenting the data. The collection contained a total of 94 photographs, 50 of which depicted PCOS, and the remaining 44 were ultrasonic photographs that did not depict PCOS.

Inan et al.~\cite{inan2021improved} suggested XGBoost for PCOS screening based on their data table. Using statistical correlation techniques such as the ANOVA Test and the Chi-Square Test, they selected  23 metabolic and clinical indicators that best identify PCOS. Using a mix of Synthetic Minority Oversampling Techniques (SMOTE) and Edited Nearest Neighbor (ENN), the data were resampled.

Lv et al.~\cite{lv2021deep} proposed a deep learning architecture employing U-net, ResNet 18, and the Multi Instance Learning model to categorize PCOS patients based on scleral pictures. There were a total of 721 photos in their dataset, of which 388 were of PCOS patients. Their proposed model attained an accuracy of 93\% and an AUC of 98\% on average.

Soni and Vashishta~\cite{soni2019image} devised an approach that employs image segmentation and convolutional neural network (CNN) to detect PCOS in ultrasound pictures. In addition, the K-Nearest Neighbor method was utilized to classify the photos.

To classify PCOS from ultrasound images, Srivastava et al.~\cite{srivastava2020detection} fine-tune the 16 Layered VGG-16 model using their own dataset. Their model achieved an accuracy of 92.11\%.

~\cite{7835370} employed image binarization on ultrasound B-mode images. Rihana et. al~\cite{rihana2013automated} converted ultrasonic pictures of ovarian cysts to grey scale images and employed image binarization as image preprocessing procedure. The authors post-processed the photos by labeling and linking components, which ultimately led to the extraction of geometrical features and classification of cysts. They utilized SVM as their classifier and attained 90\% accuracy.


\section{Methodology}
\begin{figure}[htp]
    \centering
    \includegraphics[width=8cm]{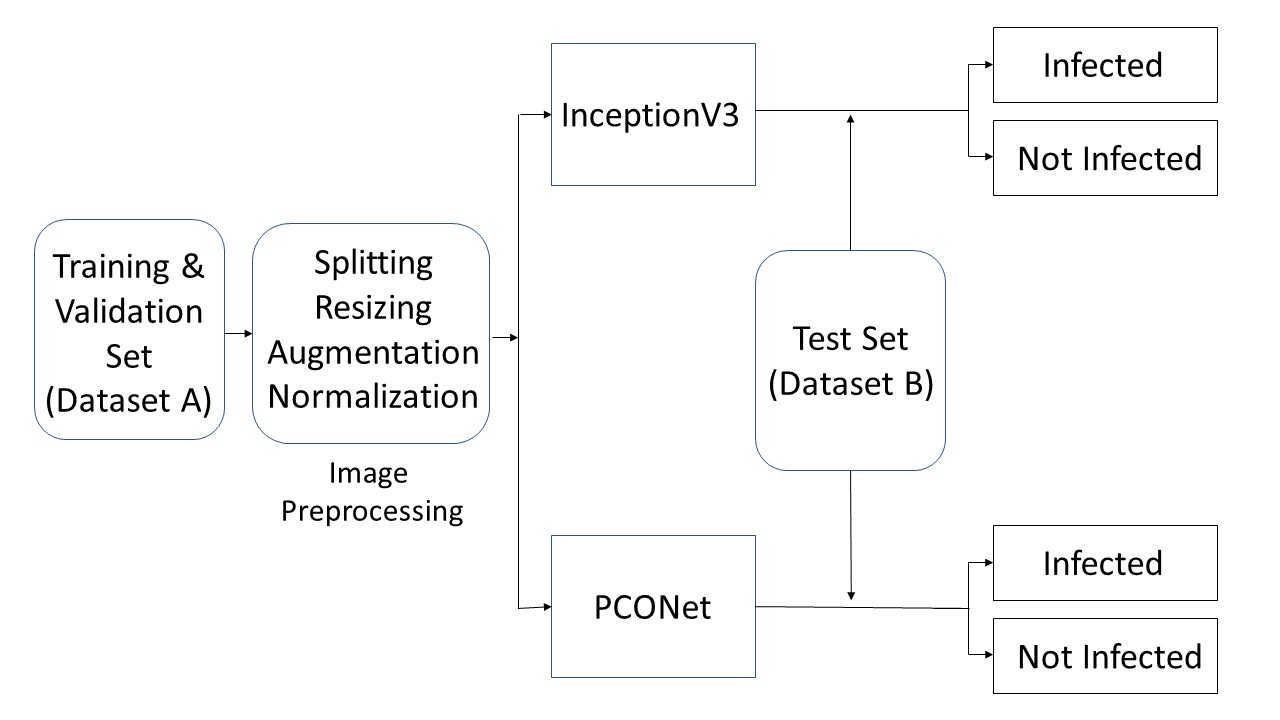}
    \caption{Workflow of training and testing our models.}
    \label{fig:methodology}
\end{figure}

Two convolutional neural network models were utilized in this study. We have fine-tuned the InceptionV3 convolutional neural network through transfer learning method and customized the model based on our dataset requirements. Additionally, we have developed a lightweight convolutional neural network to detect ovarian cysts in our dataset. Fig. \ref{fig:methodology} depicts an overview of our work process.

\subsection{Dataset Description}
\begin{figure}[htp]
    \centering
    \includegraphics[width=8cm]{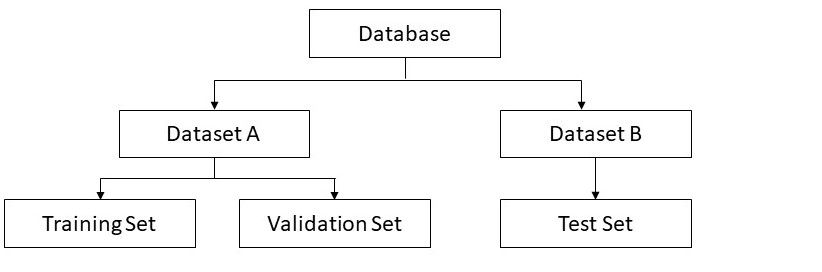}
    \caption{Overview of InceptionV3 architecture~\cite{narein}}
    \label{fig:db}
\end{figure}

In this work, our database consisted of two discrete datasets: dataset A, and dataset B. Dataset A was used for training and validating our models. But for testing, instead of partitioning a test set from dataset A, we have used a discrete dataset B for unbiased performance evaluation of the models. Fig. \ref{fig:db} depicts an overview of our whole database.

\subsubsection{Dataset A}
Our initial dataset- dataset A- which was utilized to train and validate both of our models, was obtained from Kaggle~\cite{choudhari}. Dataset A initially included a training set of 1,924 images and a test set of 1,932 images. However, the images in both the training and test sets in this dataset were largely identical. We have therefore discarded the test set and worked exclusively with the training set.

\begin{figure}[htp]
    \centering
    \includegraphics[width=8cm]{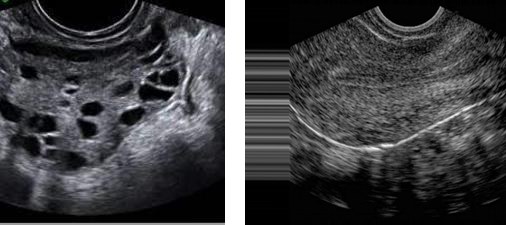}
    \caption{Overview of Dataset A. Left image is a polycystic ovarian ultrasound image and right image is a healthy ovarian ultrasound image.}
    \label{fig:dsa}
\end{figure}

We have further partitioned Dataset A's initial training set into two sets: training set and validation set. Both sets included ultrasound pictures of healthy ovaries as well as cystic ovaries. The ultrasound photos that showed infected cysts on the ovary were given the label `infected' while the ultrasound images that showed a healthy ovary were given the label `not infected'. Finally, we have come up with our final version of Dataset A which contained a total of 1,346 training photos in addition to 384 validation images. Sample images of both cystic and healthy ovarian ultrasound images are shown in Fig. \ref{fig:dsa}.

\subsubsection{Dataset B}
\begin{figure}[htp]
    \centering
    \includegraphics[width=8cm]{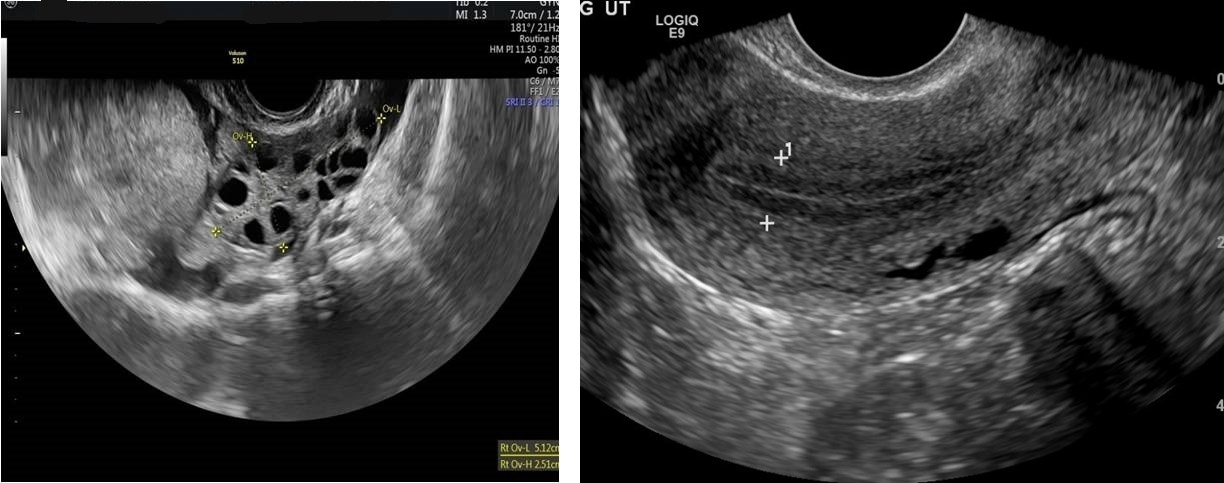}
    \caption{Overview of Dataset B. Left image is a polycystic ovarian ultrasound image and right image is a healthy ovarian ultrasound image.}
    \label{fig:dsb}
\end{figure}

For our test set- Dataset B- we have gathered ultrasound images of healthy and cystic ovarian tissues from various internet platforms. We have consulted two certified physicians in order to appropriately classify these photos as healthy or cystic. Ultrasound photos of healthy ovaries were identified as `not infected', whereas images of cystic ovaries were labeled as `infected'. Dataset B contained a total of 339 photos, 154 of which were healthy ovarian ultrasound images and 185 of which were images of ovarian cysts. This dataset was completely unrelated to dataset A to ensure an unbiased performance evaluation of our models. Sample images of both cystic and healthy ovarian ultrasound images of dataset B are shown in Fig. \ref{fig:dsb}.

\subsection{Image Preprocessing}
\label{prep}

As mentioned before, dataset A was used for training and validating our models, and dataset B for testing. Both dataset A and B included photos of varying dimensions. To provide our models with images of uniform dimensions, we have rescaled all the photos of both the datasets to 224×224 pixels. Our training, test, and validation sets each had 16 batches. We have used ImageDataGenerator from Keras to normalize our image data. We have also augmented our training images in order to overcome dataset A's image limitations. We have utilized approximately 80\% of dataset A  as training set and the remaining 20\% as validation set. Dataset B was used as test set.

\subsection{Model Description}

\subsubsection{Fine Tuning Inception-V3}

InceptionV3 was initially developed in response to the 2012 ILSVRC classification challenge~\cite{szegedy2016rethinking}. Default InceptionV3 contains 42 layers. The output layer had a dimension of 1x1x1000 because ILSVRC dataset contained 1000 classes of images, and the softmax activation function was utilized in its output layer because the goal was multiclass classification.

\begin{figure}[htp]
    \centering
    \includegraphics[width=8cm]{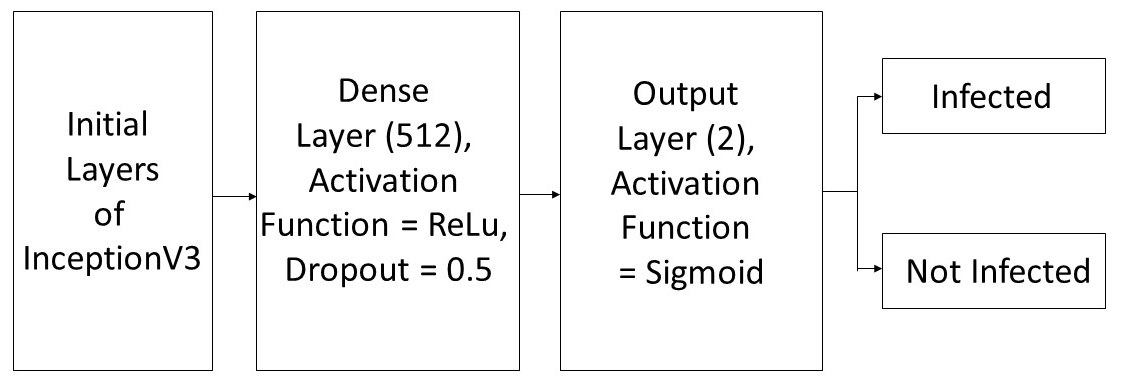}
    \caption{Overview of fine tuned InceptionV3 architecture}
    \label{fig:inceptionv3_finetuned}
\end{figure}

As our dataset required binary classification, the top layers of the InceptionV3 model were removed and two dense layers were added. The dense layer before the output layer had 512 neurons. In this layer, we have employed the ReLu activation function and a dropout of 0.5 to prevent overfitting. Our output layer consisted of two neurons, as our classification task was binary. The output layer uses Sigmoid  activation function to classify the images into two categories: `infected' and `not infected'. This model's design is depicted in Figure \ref{fig:inceptionv3_finetuned}. To compile the model, we have used Adam optimizer with a learning rate of 0.00001 and `binary crossentropy' as the loss function.

\subsubsection{PCONet}

\begin{figure}[htp]
    \centering
    \includegraphics[width=8cm]{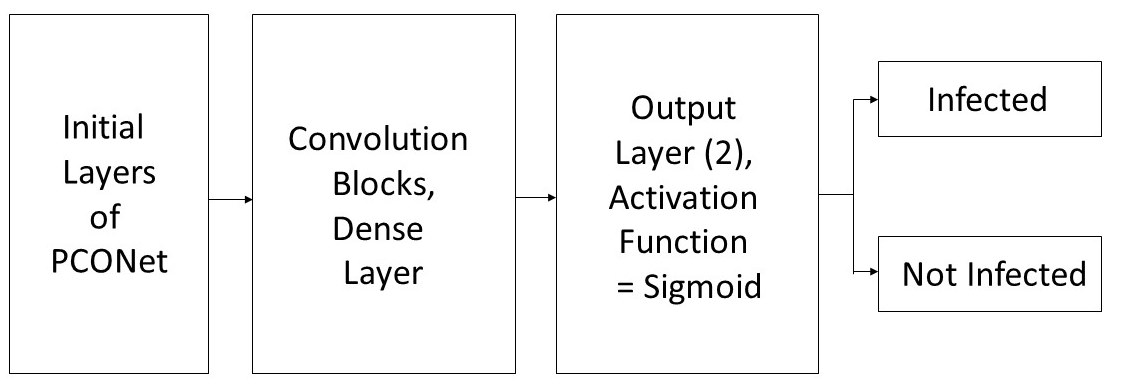}
    \caption{Simple architecture overview of PCONet.}
    \label{fig:pconet_archi}
\end{figure}

\begin{figure}[htp]
    \centering
    \includegraphics[width=8cm]{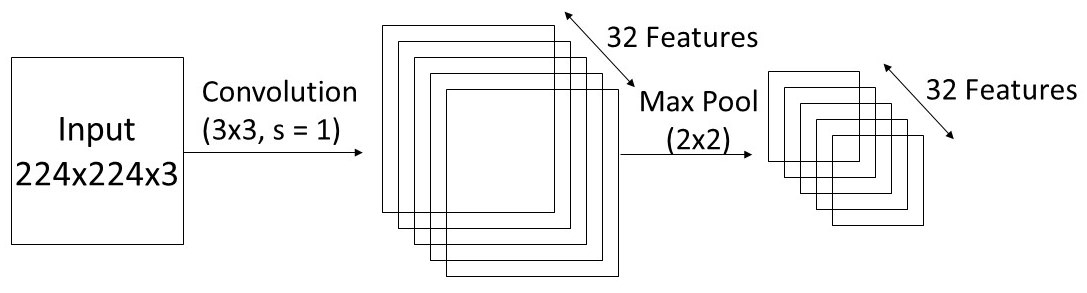}
    \caption{First Convolutional Block of PCONet.}
    \label{fig:conv_1}
\end{figure}

We have developed a convolutional neural network model to classify polycystic ovarian images. Five convolutional blocks, each consisting of a convolutional layer plus a max pooling layer, were utilized to build PCONet. The kernel size of each convolutional layer was 3x3, the activation function was ReLu, and the stride was 1. The initial convolutional layer's input shape was (224,224,3). All the maximum pooling layers had a 2x2 pool size. To extract features, the filter numbers for first through fifth convolutional layers were 32, 32, 64, 64, and 128 consecutively. Fig. \ref{fig:pconet_archi} depicts a simplified schematic of PCONet. As an example, the first convolutional block is depicted in Fig. \ref{fig:conv_1}. Such five convolutional blocks were utilized with various parameters.

Following the convolutional blocks, a flattening layer was implemented to feed data into dense layers. To classify the photos, we have utilized three dense layers. The first dense layer contained 124 neurons, a ReLu activation function, and a dropout rate of 0.5 to avoid overfitting. The second dense layer contained 256 neurons, an activation function of ReLu, and a dropout rate of 0.5. As this is a binary classification task, the output layer contained two neurons and Sigmoid  activation function. In Table \ref{table:1}, we have presented an overview of our proposed convolutional network. To compile PCONet, we have used the Adam optimizer with a learning rate of 0.00001 and the binary crossentropy loss function.

\begin{table}[]
\centering
\caption{Outline of PCONet}
\label{table:1}
\begin{tabular}{|l|l|l|}
\hline
\multicolumn{1}{|l|}{Type} & \multicolumn{1}{l|}{Specifications} & \multicolumn{1}{l|}{Output Size} \\ \hline
Conv\_1                    & 32(3, 3), s=1                       & (222, 222, 32)                   \\
Max\_pool\_1               & (2, 2), s=2                         & (111, 111, 32)                   \\
Conv\_2                    & 32(3, 3), s=1                       & (109, 109, 32)                   \\
Max\_pool\_2               & (2, 2), s=2                         & (54, 54, 32)                     \\
Conv\_3                    & 64(3, 3), s=1                       & (52, 52, 64)                     \\
Max\_pool\_3               & (2, 2), s=2                         & (26, 26, 64)                     \\
Conv\_4                    & 64(3, 3), s=1                       & (24, 24, 64)                     \\
Max\_pool\_4               & (2, 2), s=2                         & (12, 12, 64)                     \\
Conv\_5                    & 128(3, 3), s=1                      & (10, 10, 128)                    \\
Max\_pool\_5               & (2, 2), s=2                         & (5, 5, 128)                      \\
Flattening Layer           &                                     & (None, 3200)                     \\
Dense Layer 1              & 128, ReLu                           & (None, 128)                      \\
Dense Layer 2              & 256, ReLu                           & (None, 256)                      \\
Output Layer               & 2, Sigmoid                          & (None, 2)\\ \hline                 
\end{tabular}
\end{table}

\subsection{Training Models}

Our dataset included photos of various dimensions. To train both of our models, we have preprocessed all of our photos, as mentioned in section \ref{prep}. The same training set and image preprocessing procedure were followed for both models.
Both models were trained for thirty epochs. Steps per epoch were determined by dividing the number of images by the batch size. The batch size for training both models was 16. To validate the models, we have utilized our validation dataset.


\section{Result Analysis}
The quantitative performance evaluation of both our models is based on accuracy, precision, recall, and F1 Score  on training and validation sets. These images were labeled with `infected' for ploycystic ovarian ultrasound images and `not infected' for healthy ovarian images.

\begin{figure}[htp]
    \centering
    \includegraphics[width=8cm]{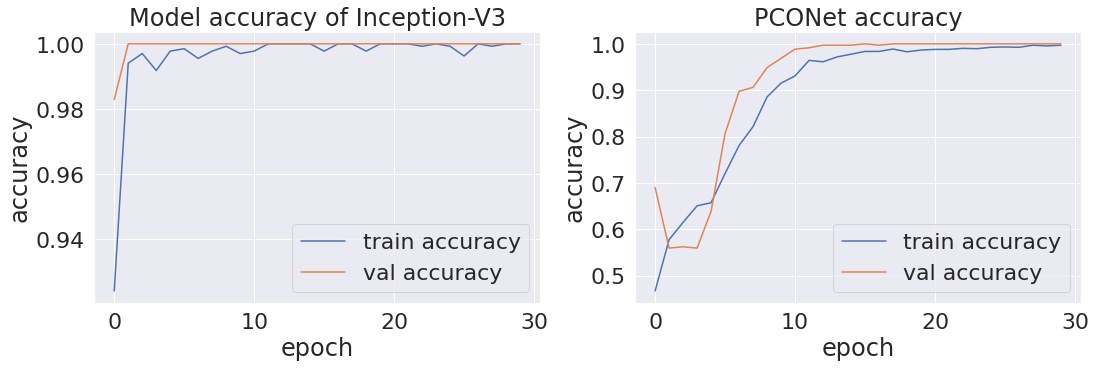}
    \caption{Training and validation accuracy of InceptionV3 (left) and PCONet (right) over 30 epochs.}
    \label{fig:acc2}
\end{figure}

\begin{figure}[htp]
    \centering
    \includegraphics[width=8cm]{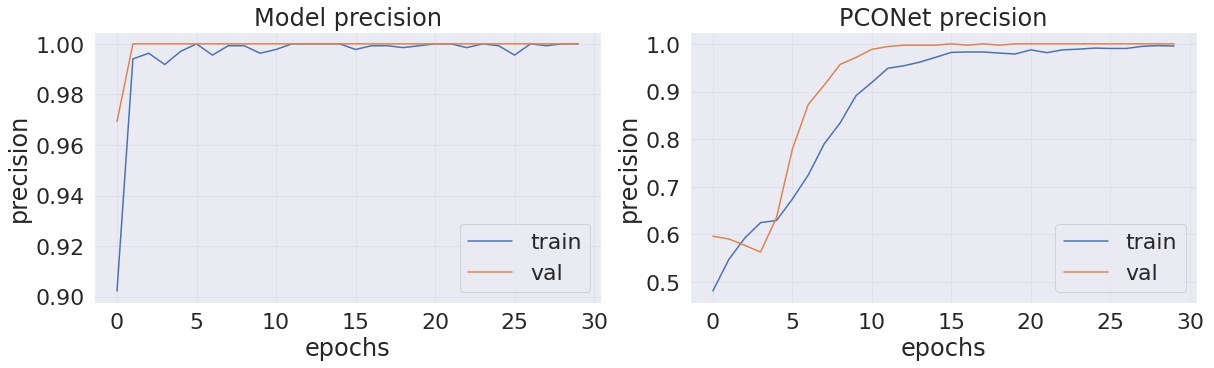}
    \caption{Training and validation precision of InceptionV3 (left) and PCONet (right) over 30 epochs.}
    \label{fig:precision}
\end{figure}

\begin{figure}[h!]
    \centering
    \includegraphics[width=8cm]{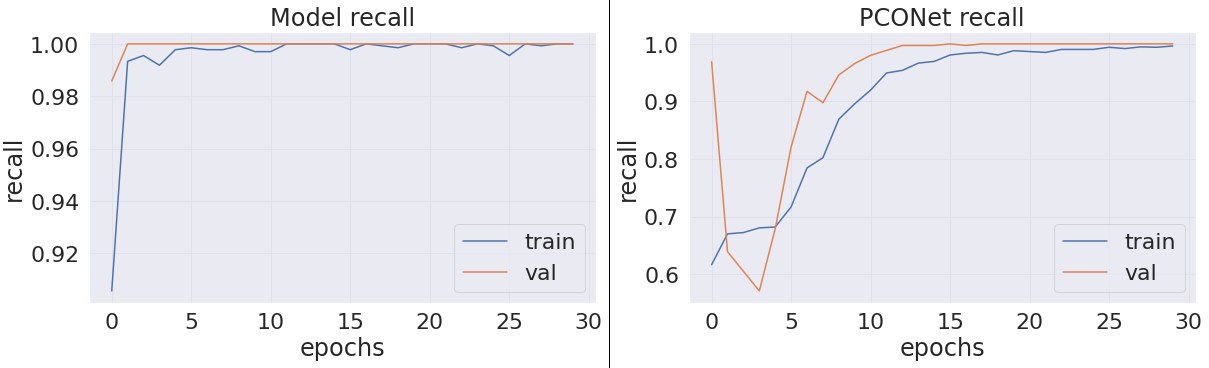}
    \caption{Training and validation recall of InceptionV3 (left) and PCONet (right) over 30 epochs}
    \label{fig:recall}
\end{figure}

\begin{figure}[htp]
    \centering
    \includegraphics[width=8cm]{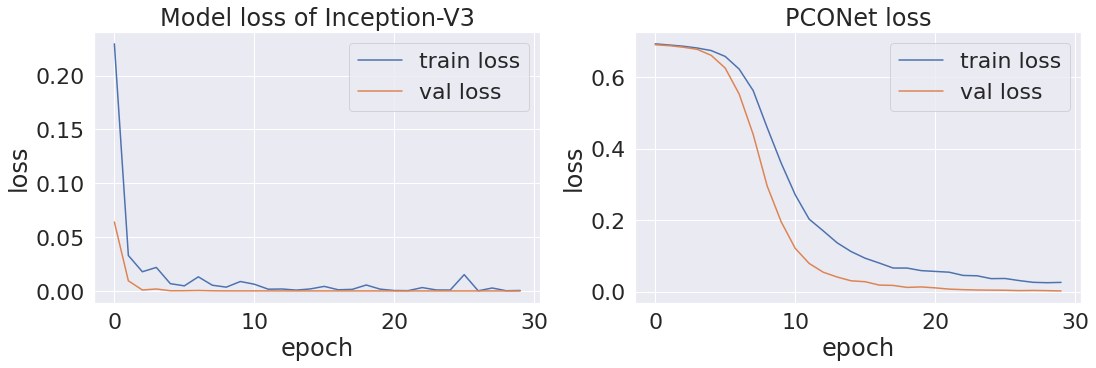}
    \caption{Training and validation loss of InceptionV3 (left) and PCONet (right) over 30 epochs}
    \label{fig:loss}
\end{figure}

PCONet showe 98.58\% accuracy and InceptionV3 showed 97.34\% accuracy in training set. PCONet showed 1.24\& higher accuracy on training set.  Confusion matrix of both the models were plotted based on this test set. A comparative table for both the models' performance on test set is given in Table \ref{table:2}. Accuracy (Fig. \ref{fig:acc2}), precision (Fig. \ref{fig:precision}), recall (Fig. \ref{fig:recall}), and loss (Fig. \ref{fig:loss}) of training and validation set were also plotted for both of the models.

\begin{table}[htp]
\centering
\caption{Comparative table of performance evaluation parameters conducted on test set for both models.}
\label{table:2}
\begin{tabular}{ |c|c|c|c| } 
 \hline
 Parameters & Class & InceptionV3 & PCONet \\
 \hline
 Accuracy & & 96.56\% & 98.12\% \\ 
 \hline
{Precision} & Cystic & 0.94 & 0.96 \\ 
 & Healthy & 1.00 & 0.97 \\ 
 \hline
{Recall} & Cystic & 1.00 & 0.97 \\
 & Healthy & 0.91 & 0.95  \\ 
 \hline
{F1 Score} & Cystic & 0.97 & 0.97 \\
 & Healthy & 0.95 & 0.96 \\ 
 \hline
\end{tabular}
\end{table}

\begin{figure}[htp]
    \centering
    \includegraphics[width=6cm]{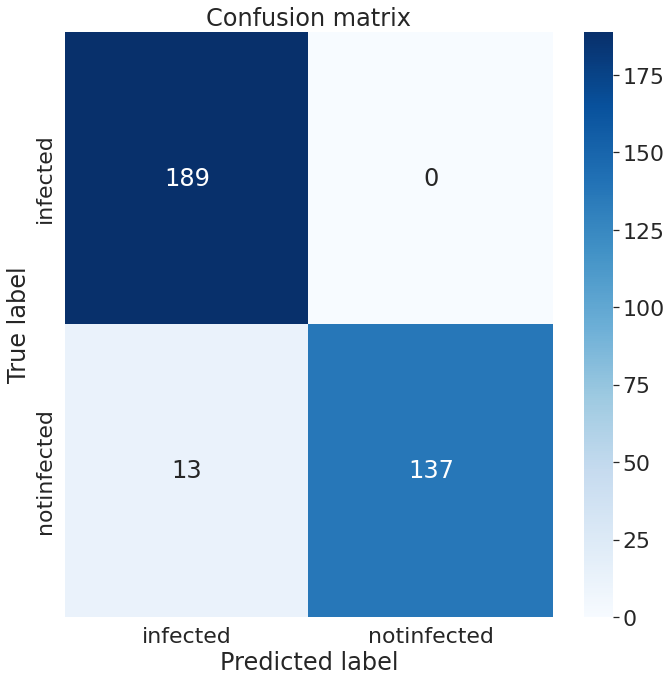}
    \caption{Confusion matrix of InceptionV3 on test set.}
    \label{fig:inception_cm}
\end{figure}

\begin{figure}[htp]
    \centering
    \includegraphics[width=6cm]{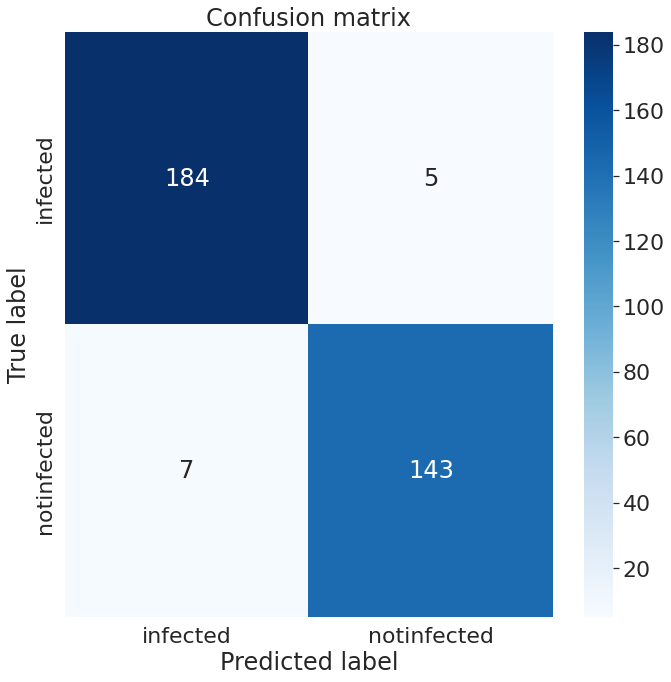}
    \caption{Confusion matrix of PCONet on test set.}
    \label{fig:pconet_cm}
\end{figure}


We have tested the accuracy of our models on dataset B, consisting of 339 images. There were 189 infected images and 150 healthy ovarian images in our test set. On our test set, InceptionV3 has an accuracy of 96.56\%, whereas PCONet has an accuracy of 98.12\%, as shown in Table \ref{table:2}. PCONet, despite being a considerably lighter model than InceptionV3, demonstrates 1.56\% higher accuracy than InceptionV3. PCONet also exhibits 2\% higher precision than InceptionV3 when it comes to the detection of cystic ovary. However, InceptionV3 demonstrates 3\% higher precision than PCONet in healthy images. In comparison to InceptionV3, PCONet had a greater recall rate for the detection of healthy ovary but a lower recall rate for the detection of cystic ovary. 

The f1 score for recognizing cystic ovary is the same for both models, which was 0.97. However, the f1-score for detecting healthy ovary is 1\% higher for the PCONet model than InceptionV3 model.

From the confusion matrix of InceptionV3 (Fig.\ref{fig:inception_cm}), and PCONet (Fig. \ref{fig:pconet_cm}), it is visible PCONet correctly predicted 5 more infected images than InceptionV3. PCONet correctly predicted all the infected images of our test set, whereas InceptionV3 had 5 incorrect infected image prediction. However, InceptionV3  correctly predicted 6 more healthy images than PCONet. InceptionV3 correctly predicted 143 healthy images out of 150, which was 137 correct predictions for PCONet.

Although the performances of both the models were almost on par with one another, PCONet showed better accuracy, which was 98.12\%, than fine tuned InceptionV3 - 96.56\%. PCONet had a total of 582,690 parameters, of which all were trainable. On the other hand, InceptionV3 had a total of 48,018,722 parameters, of which 34,432 were non-trainabkle and 47,984,290 were trainable. PCONet is a much lighter model than InceptionV3, but it still shows higher accuracy in our test dataset. As a result, we can reach the conclusion that our model PCONet is the better one among these two in detecting polycystic ovarian ultrasound images tested on our dataset.


\section{Conclusion}
In this work, we have developed PCONet- a CNN model to classify polycystic ovarian ultrasound images. We have also fine tuned pretrained InceptionV3 and compared it to PCONet on various quantitative performance evaluation parameters. To test the models, we have developed a test set completely unrelated to training and validation datasets to ensure unbiased evaluation. The PCONet showcased an accuracy of 98.12\% which was higher than the accuracy of the fine tuned InceptionV3- which showed 96.56\% accuracy. We have also compared the models on the basis of precision, recall, and f1 score. We have plotted the confusion matrix of these two models conducted on the test set. PCONet can be used to detect polycystic ovarian images in healthcare facilities to contribute to the early detection of PCOS—a vital step to mitigate the devastating effects of this syndrome. In the future, we plan to work on PCONet to further improve the accuracy. We also plan to show the elasticity of ovarian images from ultrasound~\cite{kabir2016improved}, photoacoustic ~\cite{zheng_zhang_goswami_kabir_doyley_xia_1AD} and magnetic resonance imaging ~\cite{faiyazpreliminary}.

\bibliographystyle{IEEEtran}
\bibliography{2cite}

\end{document}